\documentstyle[prd,aps,preprint,tighten,epsfig]{revtex}

\begin{document}

\draft


\title{Massive and Massless Neutrinos on Unbalanced Seesaws}
\author{{\bf Zhi-zhong Xing}
\thanks{E-mail: xingzz@ihep.ac.cn}}
\address{Institute of High Energy Physics, Chinese Academy of Sciences,
Beijing 100049, China \\ and \\
Interdisciplinary Center for Theoretical Study, University of
Science and Technology of \\ China, Hefei, Anhui 230026, China}
\maketitle

\begin{abstract}
The observation of neutrino oscillations requires new physics
beyond the standard model (SM). A SM-like gauge theory with $p$
lepton families can be extended by introducing $q$ heavy
right-handed Majorana neutrinos but preserving its $SU(2)^{}_{\rm
L} \times U(1)^{}_{\rm Y}$ gauge symmetry. The overall neutrino
mass matrix $M$ turns out to be a symmetric $(p+q) \times (p+q)$
matrix. Given $p>q$, the rank of $M$ is in general equal to $2q$,
corresponding to $2q$ non-zero mass eigenvalues. The existence of
$(p-q)$ massless left-handed Majorana neutrinos is an exact
consequence of the model, independent of the usual approximation
made in deriving the Type-I seesaw relation between the effective
$p \times p$ light Majorana neutrino mass matrix $M^{}_\nu$ and
the $q \times q$ heavy Majorana neutrino mass matrix $M^{}_{\rm
R}$. In other words, the numbers of {\it massive} left- and
right-handed neutrinos are fairly matched. A good example to
illustrate this ``seesaw fair play rule" is the minimal seesaw
model with $p=3$ and $q=2$, in which one massless neutrino sits on
the {\it unbalanced} seesaw.
\end{abstract}


\newpage

Very robust evidence for non-zero neutrino masses and large lepton
flavor mixing has recently been achieved from solar \cite{SNO},
atmospheric \cite{SK}, reactor \cite{KM} and accelerator
\cite{K2K} neutrino oscillation experiments. This great
breakthrough opens a new window to physics beyond the standard
model (SM). So far a number of theoretical scenarios have been
proposed, either at low energy scales or at high energy scales, to
understand why the masses of neutrinos are considerably smaller
than those of charged leptons and quarks \cite{Review}. Among
them, the seesaw mechanism \cite{SS} seems to be most elegant and
natural. In particular, an appropriate combination of the seesaw
mechanism and the leptogenesis mechanism \cite{FY} allows one to
simultaneously account for the observed neutrino oscillations and
the observed matter-antimatter asymmetry of the Universe.

\vspace{0.2cm}

The canonical (Type-I) seesaw idea is rather simple, indeed. By
introducing three right-handed Majorana neutrinos to the SM and
keeping its Lagrangian invariant under the $SU(2)^{}_{\rm L}
\times U(1)^{}_{\rm Y}$ gauge transformation, one may write out a
normal Dirac neutrino mass term relevant to the electroweak
symmetry breaking ($M^{}_{\rm D}$) and an extra Majorana neutrino
mass term irrelevant to the electroweak symmetry breaking
($M^{}_{\rm R}$). Given $M^{}_{\rm D}$ as the seesaw fulcrum at or
close to the electroweak symmetry breaking scale ($\sim 10^2$
GeV), the smallness of three left-handed neutrino masses ($< 1$
eV) is then attributed to the largeness of three right-handed
neutrino masses ($> 10^{13}$ GeV) \cite{SS}: $M^{}_\nu \approx
-M^{}_{\rm D} M^{-1}_{\rm R} M^T_{\rm D}$. Since both $M^{}_{\rm
D}$ and $M^{}_{\rm R}$ are in general the rank-3 matrices,
$M^{}_\nu$ is also of rank 3 and thus has three non-vanishing mass
eigenvalues.

\vspace{0.2cm}

Can massive and massless neutrinos coexist in a general seesaw
scenario? Such a question makes sense for two simple reasons. On
the one hand, current neutrino oscillation data do allow one of
the light neutrinos to be massless or almost massless (e.g.,
either $m^{}_1 \rightarrow 0$ or $m^{}_3 \rightarrow 0$
\cite{FGY}). On the other hand, it is conceptually interesting to
distinguish between the neutrino with an exact zero mass and the
neutrino with a vanishingly small mass. An affirmative answer to
the above question has been observed in Refs. \cite{Valle,Liao}.
The purpose of this short note is to have a new look at the
properties of massive and massless neutrinos in the generalized
Type-I seesaw mechanism. We shall consider a straightforward
extension of the SM with $p$ lepton families, $q$ heavy
right-handed Majorana neutrinos and the $SU(2)^{}_{\rm L} \times
U(1)^{}_{\rm Y}$ gauge symmetry. The overall neutrino mass matrix
$M$ in this model turns out to be a symmetric $(p+q) \times (p+q)$
matrix. Given $p>q$, the rank of $M$ is in general equal to $2q$,
corresponding to $2q$ non-zero mass eigenvalues. We demonstrate
that the existence of $(p-q)$ massless left-handed Majorana
neutrinos is an exact consequence of the model, independent of the
usual approximation made in deriving the Type-I seesaw relation
between the effective $p \times p$ light Majorana neutrino mass
matrix $M^{}_\nu$ and the $q \times q$ heavy Majorana neutrino
mass matrix $M^{}_{\rm R}$. We refer to this kind of seesaw, in
which the number of left-handed neutrinos is larger than the
number of right-handed neutrinos, as the {\it unbalanced} seesaw.
The fact that the numbers of {\it massive} left- and right-handed
Majorana neutrinos are fairly matched on unbalanced seesaws can be
referred to as the ``seesaw fair play rule". A well-known example
is the minimal seesaw model with $p=3$ and $q=2$ \cite{FGY}, in
which one massless neutrino sits on the unbalanced seesaw. The
stability of $m^{}_i =0$ against radiative corrections from the
seesaw scale down to the electroweak scale will also be stressed.

\vspace{0.5cm}

Let us consider a simple extension of the SM with $p$ lepton
families and $q$ heavy right-handed Majorana neutrinos. The
Lagrangian of this electroweak model is required to be invariant
under the $SU(2)^{}_{\rm L} \times U(1)^{}_{\rm Y}$ gauge
transformation. To be explicit, the lepton mass terms can be
written as
\begin{equation}
-{\cal L}^{}_{\rm lepton} \; =\; \overline{l^{}_{\rm L}} Y^{}_l
e^{}_{\rm R} H ~ + ~ \overline{l^{}_{\rm L}} Y^{}_\nu N^{}_{\rm R}
H^{\rm c} ~ + ~ \frac{1}{2} \overline{N^{\rm c}_{\rm R}} M^{}_{\rm
R} N^{}_{\rm R} ~ + ~ {\rm h.c.} \; ,
\end{equation}
where $l^{}_{\rm L}$ denotes the left-handed lepton doublets;
$e^{}_{\rm R}$ and $N^{}_{\rm R}$ stand respectively for the
right-handed charged-lepton and Majorana neutrino singlets; $H$ is
the Higgs-boson weak isodoublet (with $H^{\rm c} \equiv
i\sigma^{}_2 H^*$); $M^{}_{\rm R}$ is the $q \times q$ heavy
Majorana neutrino mass matrix; $Y^{}_l$ and $Y^{}_\nu$ are the
coupling matrices of charged-lepton and neutrino Yukawa
interactions. After spontaneous gauge symmetry breaking, the
neutral component of $H$ acquires the vacuum expectation value $v
\approx 174$ GeV. Then we arrive at the $p \times p$
charged-lepton mass matrix $M^{}_l = v Y^{}_l$ and the $p \times
q$ Dirac neutrino mass matrix $M^{}_{\rm D} = v Y^{}_\nu$. Eq. (1)
turns out to be
\begin{equation}
-{\cal L}^\prime_{\rm lepton} \; =\; \overline{e^{}_{\rm L}}
M^{}_l e^{}_{\rm R} ~ + ~ \frac{1}{2} ~\overline{(\nu^{}_{\rm L}
~~~ N^{\rm c}_{\rm R} )} \left ( \matrix{ {\bf 0}     & M^{}_{\rm
D} \cr\cr M^T_{\rm D}    & M^{}_{\rm R} \cr} \right ) \left (
\matrix{ \nu^{\rm c}_{\rm L} \cr\cr N^{}_{\rm R} \cr} \right ) ~ +
~ {\rm h.c.} \; ,
\end{equation}
where $e$, $\nu^{}_{\rm L}$ and $N^{}_{\rm R}$ represent the
column vectors of $p$ charged-lepton fields, $p$ left-handed
neutrino fields and $q$ right-handed neutrino fields,
respectively. In obtaining Eq. (2), we have made use of the
relation $\overline{\nu^{}_{\rm L}} M^{}_{\rm D} N^{}_{\rm R} =
\overline{N^{\rm c}_{\rm R}} M^T_{\rm D} \nu^{\rm c}_{\rm L}$ as
well as the properties of $\nu^{}_{\rm L}$ (or $N^{}_{\rm R}$) and
$\nu^{\rm c}_{\rm L}$ (or $N^{\rm c}_{\rm R}$) \cite{Xing04}. Note
that the mass scale of $M^{}_{\rm R}$ can naturally be much higher
than the electroweak scale $v$, because those right-handed
Majorana neutrinos are $SU(2)^{}_{\rm L}$ singlets and their
corresponding mass term is not subject to the magnitude of $v$.
The overall neutrino mass matrix
\begin{equation}
M \; =\; \left ( \matrix{ {\bf 0}     & M^{}_{\rm D} \cr\cr
M^T_{\rm D}    & M^{}_{\rm R} \cr} \right )
\end{equation}
is a symmetric $(p+q) \times (p+q)$ matrix and can be diagonalized
by the transformation
\begin{equation}
U^\dagger M U^* \; =\; \left( \matrix{ m^{}_1 & & & & & \cr &
\ddots & & & & \cr & & m^{}_p & & & \cr & & & M^{}_1 & & \cr & & &
& \ddots & \cr & & & & & M^{}_q \cr} \right) \; ,
\end{equation}
where $U$ is a unitary matrix, $m^{}_i$ (for $i = 1, \cdots, p$)
denote the masses of $p$ left-handed Majorana neutrinos, and
$M^{}_j$ (for $j = 1, \cdots, q$) denote the masses of $q$
right-handed Majorana neutrinos. If the mass scale of $M^{}_{\rm
R}$ is considerably higher than that of $M^{}_{\rm D}$, one may
obtain the {\it effective} light neutrino mass matrix
\begin{equation}
M^{}_\nu \; \approx \; -M^{}_{\rm D} M^{-1}_{\rm R} M^T_{\rm D}
\end{equation}
as an extremely good approximation \cite{Zhou}. In this Type-I
seesaw scenario, the mass eigenvalues of $M^{}_\nu$ and $M^{}_{\rm
R}$ are $m^{}_i$ (for $i = 1, \cdots, p$) and $M^{}_j$ (for $j =
1, \cdots, q$), respectively, to a high degree of accuracy. Of
course, $m^{}_i \ll v$ and  $M^{}_j \gg v$ naturally hold. Our
concern is whether some of $m^{}_i$ can in general be vanishing.

\vspace{0.2cm}

We focus on the $p>q$ case, since the $p<q$ case is less motivated
from the viewpoint of maximum simplicity and predictability in
building a seesaw model and interpreting the experimental data.
Given $p>q$, the rank of $M^{}_\nu$ is determined by that of
$M^{}_{\rm R}$ through the seesaw relation $M^{}_\nu \approx
-M^{}_{\rm D} M^{-1}_{\rm R} M^T_{\rm D}$. Namely, $M^{}_\nu$ must
be of rank $q$ in general
\footnote{Here ``in general" means that any contrived textures of
$M^{}_{\rm D}$, which might reduce the rank of $M^{}_\nu$ from $q$
to a smaller integer, are not taken into account. Without loss of
generality, $M^{}_{\rm R}$ can always be taken to be diagonal and
positive. In this basis, a too special texture of $M^{}_{\rm D}$
is usually disfavored in order to simultaneously account for
current neutrino oscillation data and the cosmological baryon
number asymmetry \cite{Review,FGY}.}.
Because the number of non-zero eigenvalues of a symmetric matrix
is equal to the rank of this matrix \cite{Rank}, we can conclude
that $M^{}_\nu$ has $(p-q)$ vanishing mass eigenvalues. Note that
this statement relies on the Type-I seesaw relation which directly
links $M^{}_{\rm R}$ to $M^{}_\nu$. Taking account of the
approximation made in deriving this seesaw formula (no matter how
good it is), we have to clarify whether the $(p-q)$ mass
eigenvalues of $M^{}_\nu$ are exactly vanishing or only
vanishingly small. A reliable proof or disproof of the above
statement should be independent of the approximate seesaw
relation.

\vspace{0.2cm}

So what we need to do is to calculate the rank of $M$ in Eq. (3).
Taking
\begin{eqnarray}
M^{}_{\rm D} & = & \left( \matrix{D^{}_{11} & \cdots & D^{}_{1q}
\cr \vdots & \ddots & \vdots \cr D^{}_{p1} & \cdots & D^{}_{pq}
\cr} \right) \; , \nonumber \\ \nonumber \\
M^{}_{\rm R} & = & \left( \matrix{R^{}_{11} & \cdots & R^{}_{1q}
\cr \vdots & \ddots & \vdots \cr R^{}_{q1} & \cdots & R^{}_{qq}
\cr} \right) \; ,
\end{eqnarray}
where $R^{}_{ij} = R^{}_{ji}$ (for $i, j =1, \cdots, q$), we write
out the explicit expression of $M$:
\begin{equation}
M \; =\; \left( \matrix{0 & \cdots & 0 & D^{}_{11} & \cdots &
D^{}_{1q} \cr \vdots & \ddots & \vdots & \vdots & \ddots & \vdots
\cr 0 & \cdots & 0 & D^{}_{p1} & \cdots & D^{}_{pq} \cr\cr
D^{}_{11} & \cdots & D^{}_{p1} & R^{}_{11} & \cdots & R^{}_{1q}
\cr \vdots & \ddots & \vdots & \vdots & \ddots & \vdots \cr
D^{}_{1q} & \cdots & D^{}_{pq} & R^{}_{q1} & \cdots & R^{}_{qq}
\cr} \right) \; .
\end{equation}
By definition, the rank of $M$ is the number of non-zero rows in
the reduced row echelon form of $M$. The latter can be calculated
by using the method of Gauss elimination. Because the upper-left
$p\times p$ sub-matrix is a zero matrix, it is easy to convert the
upper-right $p\times q$ sub-matrix (i.e., $M^{}_{\rm D}$) into a
reduced row echelon form in which the first $(p-q)$ rows are full
of zero elements. In contrast, the lower-right $q\times q$
sub-matrix (i.e., $M^{}_{\rm R}$) is of rank $q$. The rank of $M$
turns out to be $p - (p-q) + q = 2q$, corresponding to $2q$
non-zero mass eigenvalues. In other words, $q$ of the $p$ light
Majorana neutrinos must be massive, and the remaining $(p-q)$
light Majorana neutrinos must be exactly massless.

\vspace{0.2cm}

If a seesaw scenario includes unequal numbers of light
(left-handed) and heavy (right-handed) Majorana neutrinos, it can
be referred to as an {\it unbalanced} seesaw scenario. When the
number of light neutrinos is larger than that of heavy neutrinos,
such an unbalanced seesaw is actually balanced because all the
redundant light neutrinos are massless. That is, the number of
{\it massive} left-handed Majorana neutrinos is fairly equal to
the number of heavy right-handed Majorana neutrinos. We refer to
this interesting observation, which is independent of the
approximation made in deriving the Type-I seesaw formula, as the
``seesaw fair play rule" (see FIG. 1 for illustration). One can
see later on that such a rule is not only conceptually appealing
but also applicable to an instructive and
phenomenologically-favored model, the minimal seesaw model
\cite{FGY}.

\vspace{0.5cm}

To be realistic, one has to fix $p=3$ for the number of
left-handed neutrinos. Then only $q=1$ and $q=2$ are of interest
for the discussion of unbalanced seesaw scenarios. The $q=1$ case
is not favored in the Type-I seesaw framework, because it requires
two left-handed Majorana neutrinos to be massless and thus cannot
accommodate two independent neutrino mass-squared differences
observed in solar and atmospheric neutrino oscillations (i.e.,
$\Delta m^2_{21} = m^2_2 - m^2_1 \approx 8 \times 10^{-5} ~ {\rm
eV}^2$ and $\Delta m^2_{32} = m^2_3 - m^2_2 \approx \pm 2.5 \times
10^{-3} ~ {\rm eV}^2$ \cite{Vissani}). On the other hand, the
$q=2$ case is compatible with current experimental data and has
been referred to as the minimal seesaw model \cite{FGY} for the
study of both neutrino mixing and baryogenesis via leptogenesis.

\vspace{0.2cm}

According to the ``seesaw fair play rule", there must exist one
massless neutrino in the minimal seesaw model. One may also get at
this point by calculating the determinant of the $5\times 5$
neutrino mass matrix $M$, in which the Dirac neutrino mass matrix
$M^{}_{\rm D}$ is $3\times 2$ and the right-handed Majorana
neutrino mass matrix $M^{}_{\rm R}$ is $2\times 2$. It is very
straightforward to prove ${\rm Det}M =0$. Since $\left |{\rm Det}M
\right | = m^{}_1 m^{}_2 m^{}_3 M^{}_1 M^{}_2$ holds, one of
$m^{}_i$ (for $i=1,2,3$) must be vanishing. The solar neutrino
oscillation experiment has fixed $m^{}_2 > m^{}_1$ \cite{Vissani},
and thus we are left with two distinct possibilities:
\begin{itemize}
\item       $m^{}_1 = 0$, corresponding to a normal neutrino mass
hierarchy. Taking account of current experimental data, we can
easily obtain $m^{}_2 = \sqrt{\Delta m^2_{21}} ~ \approx 8.9
\times 10^{-3} ~ {\rm eV}$ and $m^{}_3 = \sqrt{\Delta m^2_{21} +
|\Delta m^2_{32}|} ~ \approx 5.1 \times 10^{-2} ~ {\rm eV}$.

\item       $m^{}_3 = 0$, corresponding to an inverted neutrino mass
hierarchy. Taking account of current experimental data, we arrive
at $m^{}_1 = \sqrt{|\Delta m^2_{32}| - \Delta m^2_{21}} ~ \approx
4.9 \times 10^{-2} ~ {\rm eV}$ and $m^{}_2 = \sqrt{|\Delta
m^2_{32}|} ~ \approx 5.0 \times 10^{-2} ~ {\rm eV}$.
\end{itemize}
Note that it is possible to build viable neutrino models
\cite{Example} to accommodate both a special neutrino mass
spectrum with $m^{}_1 =0$ or $m^{}_3 =0$ and the (nearly)
tri-bimaximal neutrino mixing pattern \cite{TB}. Some of such
models can even provide a natural interpretation of the
cosmological baryon number asymmetry via (resonant) leptogenesis.

\vspace{0.2cm}

It is worth mentioning that $m^{}_1 =0$ (or $m^{}_3 =0$) is stable
against radiative corrections from the seesaw scale (usually
measured by the lightest right-handed Majorana neutrino mass
$M^{}_1$) down to the electroweak scale (usually characterized by
the $Z^0$ mass $M^{}_Z$ or simply the vacuum expectation value of
the neutral Higgs field $v$) in the minimal seesaw model, at least
at the one-loop level \cite{Mei}. This observation is also
expected to be true for a general unbalanced seesaw scenario with
$p>q$; namely, the zero masses of left-handed Majorana neutrinos
in such a scenario are insensitive to radiative corrections
between the scales $M^{}_Z$ and $M^{}_1$. Therefore, it makes
sense to study the phenomenology of unbalanced seesaw models in
which massive and massless neutrinos coexist.

\vspace{0.5cm}

To summarize, we have considered a SM-like $SU(2)^{}_{\rm L}
\times U(1)^{}_{\rm Y}$ gauge theory with $p$ lepton families and
$q$ heavy right-handed Majorana neutrinos. Given $p>q$, we have
shown that the overall $(p+q) \times (p+q)$ neutrino mass matrix
$M$ is in general of rank $2q$, corresponding to $2q$ non-zero
mass eigenvalues. An important emphasis is that the existence of
$(p-q)$ massless left-handed Majorana neutrinos is an exact
consequence of the model, independent of the usual approximation
made in deriving the Type-I seesaw relation between the effective
$p \times p$ light Majorana neutrino mass matrix $M^{}_\nu$ and
the $q \times q$ heavy Majorana neutrino mass matrix $M^{}_{\rm
R}$. In other words, the numbers of {\it massive} left- and
right-handed neutrinos are fairly matched in such an unbalanced
seesaw scenario. We have taken the minimal seesaw model (with
$p=3$ and $q=2$) as a simple but realistic example, in which one
massless left-handed neutrino coexists with two massive
left-handed neutrinos, to illustrate this ``seesaw fair play
rule".

\vspace{0.2cm}

Since the seesaw mechanism is a particularly natural, concise and
appealing mechanism to understand the smallness of left-handed
Majorana neutrino masses, its potential properties deserve further
investigation. The main point of this note is that massless and
massive neutrinos can coexist in an unbalanced seesaw scenario, if
the number of heavy right-handed Majorana neutrinos is smaller
than that of light left-handed Majorana neutrinos. Whether one of
the light neutrinos is really massless or not remains an open
question, but it is certainly a meaningful question and should be
answered experimentally in the future. On the theoretical side, it
is also of interest to explore a complete seesaw picture for
neutrino mass generation, lepton flavor mixing, CP violation and
leptogenesis with mismatched numbers of light and heavy Majorana
neutrinos.

\vspace{0.5cm}

I would like to thank D.N. Gao and J.X. Lu for warm hospitality at
the Interdisciplinary Center for Theoretical Study of USTC, where
this note was written. I am also indebted to Y. Liao for drawing
my attention to Refs. \cite{Valle,Liao}. This work was supported
in part by the National Natural Science Foundation of China.

\newpage

\begin{figure}[t]
\vspace{-9cm}
\epsfig{file=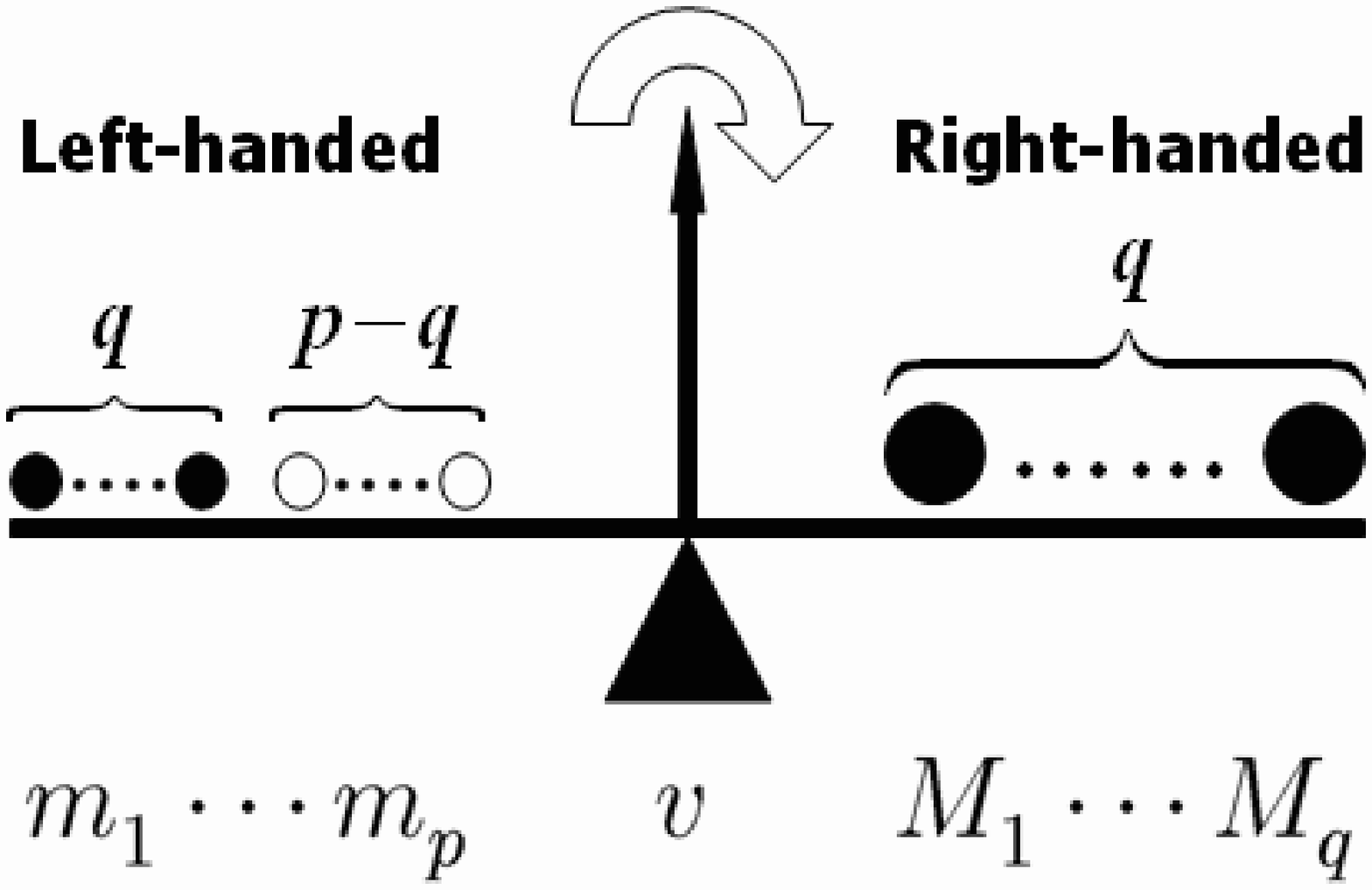,bbllx=1.5cm,bblly=14cm,bburx=13.8cm,bbury=34cm,%
width=10cm,height=16cm,angle=0,clip=0} \vspace{7.5cm}
\caption{Illustration of the ``seesaw fair play rule": if the
number of left-handed Majorana neutrinos ($p$) is larger than that
of right-handed Majorana neutrinos ($q$) on an unbalanced seesaw,
all the redundant light neutrinos ($p-q$) must be massless. In
other words, the number of massive left-handed neutrinos is fairly
equal to that of heavy right-handed neutrinos in this seesaw
scenario.}
\end{figure}


\begin{thebibliography}{99}
 \bibitem{SNO} SNO Collaboration, Q.R. Ahmad {\it et al.},
Phys. Rev. Lett. {\bf 89}, 011301 (2002).

\bibitem{SK} For a review, see: C.K. Jung {\it et al.},
Ann. Rev. Nucl. Part. Sci. {\bf 51}, 451 (2001).

\bibitem{KM} KamLAND Collaboration, K. Eguchi {\it et al.},
Phys. Rev. Lett. {\bf 90}, 021802 (2003); CHOOZ Collaboration, M.
Apollonio {\it et al.}, Phys. Lett. B {\bf 420}, 397 (1998); Palo
Verde Collaboration, F. Boehm {\it et al.}, Phys. Rev. Lett. {\bf
84}, 3764 (2000).

\bibitem{K2K} K2K Collaboration, M.H. Ahn {\it et al.},
Phys. Rev. Lett. {\bf 90}, 041801 (2003).

\bibitem{Review} For recent reviews with extensive references, see, e.g.,
H. Fritzsch and Z.Z. Xing, Prog. Part. Nucl. Phys. {\bf 45}, 1
(2000); G. Altarelli and F. Feruglio, New J. Phys. {\bf 6}, 106
(2004); R.N. Mohapatra and A.Yu. Smirnov, Ann. Rev. Nucl. Part.
Sci. {\bf 56}, 569 (2006).

\bibitem{SS} P. Minkowski, Phys. Lett. B {\bf 67}, 421 (1977);
T. Yanagida, in {\it Proceedings of the Workshop on Unified Theory
and the Baryon Number of the Universe}, edited by O. Sawada and A.
Sugamoto (KEK, Tsukuba, 1979), p. 95; M. Gell-Mann, P. Ramond, and
R. Slansky, in {\it Supergravity}, edited by F. van Nieuwenhuizen
and D. Freedman (North Holland, Amsterdam, 1979), p. 315; S.L.
Glashow, in {\it Quarks and Leptons}, edited by M.
L$\rm\acute{e}vy$ {\it et al.} (Plenum, New York, 1980), p. 707;
R.N. Mohapatra and G. Senjanovic, Phys. Rev. Lett. {\bf 44}, 912
(1980).

\bibitem{FY} M. Fukugita and T. Yanagida, Phys. Lett. B {\bf 174},
45 (1986).

\bibitem{FGY} P. Frampton, S.L. Glashow, and T. Yanagida, Phys.
Lett. B {\bf 548}, 119 (2002). For a recent review with extensive
references, see: W.L. Guo, Z.Z. Xing, and S. Zhou, Int. J. Mod.
Phys. E {\bf 16}, 1 (2007).

\bibitem{Valle} J. Schechter and J.W.F. Valle, Phys. Rev. D {\bf
22}, 2227 (1980).

\bibitem{Liao} Y. Liao, Nucl. Phys. B {\bf 749}, 153 (2006);
Eur. Phys. J. C {\bf 49}, 783 (2007).

\bibitem{Xing04} See, e.g., Z.Z. Xing, Int. J. Mod. Phys. A {\bf 19}, 1
(2004).

\bibitem{Zhou} For a detailed discussion about the approximation made
in deriving the Type-I seesaw relation, see: Z.Z. Xing and S.
Zhou, High Energy Phys. Nucl. Phys. {\bf 30}, 828 (2006);
hep-ph/0512290.

\bibitem{Rank} See, e.g., R.A. Horn and C.R. Johnson, {\it Matrix
Analysis} (Cambridge University Press, 1985).

\bibitem{Vissani} See, e.g., A. Strumia and F. Vissani,
hep-ph/0606054.

\bibitem{Example} S. Chang, S.K. Kang, and K. Siyeon,
Phys. Lett. B {\bf 597}, 78 (2004); Z.Z. Xing and S. Zhou,
hep-ph/0607302; S. Luo and Z.Z. Xing, Phys. Lett. B {\bf 646}, 242
(2007); T. Kitabayashi, hep-ph/0703303; A.H. Chan, H. Fritzsch,
and Z.Z. Xing, arXiv:0704.3153 [hep-ph]; R. Friedberg and T.D.
Lee, arXiv:0705.4156 [hep-ph].

\bibitem{TB} P.F. Harrison, D.H. Perkins, and W.G. Scott, Phys.
Lett. B {\bf 530}, 167 (2002); Z.Z. Xing, Phys. Lett. B {\bf 533},
85 (2002); P.F. Harrison and W.G. Scott, Phys. Lett. B {\bf 535},
163 (2002); X.G. He and A. Zee, Phys. Lett. B {\bf 560}, 87
(2003).

\bibitem{Mei} J.W. Mei and Z.Z. Xing, Phys. Rev. D {\bf 69}, 073003
(2004); S. Luo and Z.Z. Xing, Phys. Lett. B {\bf 637}, 279 (2006).

\end{thebibliography}
\end{document}